Version: 2025

# Confidences in hypotheses

by

Graham N. Bornholt [1]

***Summary*:** This article extends the hypotheses assessment method to the case with two competing simple hypotheses. In doing so we further clarify the benefits that hypotheses assessments can bring to classical statistical analyses. Given that confidences in hypotheses are based on conditional probabilities, we address the issue of what to condition on in order to avoid poor conditional properties. This step is essential if the resulting inferences are to be relevant to the data at hand. Admissibility is addressed within a framework of seeking confidences that are relevant to the data at hand and are as powerful as the application allows. Confidence procedures are said to be consistent if they are free of super-relevant betting strategies. For simple hypotheses, the assessment method produces minimum and maximum confidences in each hypothesis. Assessments for both symmetric and asymmetric experiments are included, and the relationship with Bayesian posterior probabilities discussed.

[1] Address: Kingscliff NSW 2487, Australia. Email: grahambornholt@gmail.com.

## 1. INTRODUCTION

In situations with two competing hypotheses, hypotheses assessment is the frequentist procedure designed to answer the question: "Given the sample data (and assumed model), what are the frequentist confidence levels for each hypothesis?" (Bornholt, 2025, p.1.) Bornholt addressed this question for the case with composite one-sided location hypotheses. Confidences in hypotheses are the frequentist analogues to the corresponding Bayesian posterior probabilities. In this paper we extend the assessment method to cover situations with two competing *simple* hypotheses, say $H_0$ and $H_1$.

Such confidences can be used on a standalone basis, or they can be used to complement test outcomes. The latter approach has clear benefits for researchers for the following reason. A $p$-value can be regarded as providing information on the *absolute* plausibility of the null hypothesis because it is a measure of the compatibility of the data with just the null hypothesis. In contrast, assessments are frequentist measures of the *relative* plausibility of each hypothesis. For instance, in Example 3 a $p$-value of 0.0197 coincides with at least 92% confidence in $H_1$ and at most 8% confidence in $H_0$. In such cases, having both the test outcome and the assessments available can leave applied researchers more informed about the evidence provided by the data.

Such confidences are derived from probabilities. The aim is to produce frequentist confidences in the hypotheses that are relevant to the data at hand and that are as powerful as the particular application allows. Following Bornholt (2025), in Section 2 confidences in the hypotheses are based on conditional probabilities that the maximum likelihood hypothesis and minimum hypothesis estimators select the correct hypothesis. (Assessment confidences are based on conditional rather than unconditional probabilities to ensure that the resulting confidences will be sufficiently relevant to the data at hand.) Example 1 shows that confidence levels can be too-high or too-low from a post-sample relevance perspective if care is not taken pre-sample to avoid procedures with poor conditional properties. Thus, how to avoid procedures with poor conditional properties becomes a key question.

In Section 3, we address this issue and select the weakest of Robinson's (1979) set of definitions of desirable conditional properties as a consistency requirement that confidence claims should satisfy. (As a byproduct of this requirement, in some cases confidence statements may be more conservative than could otherwise be achieved.) Requiring that assessments take the form of consistent confidences ensures both frequentist interpretability and relevance to the data at hand. There remains the issue of what variable to condition on (or, equivalently, how to best partition the sample space). In Section 4, we address this question using an optimization that formalizes mathematically the aim to produce frequentist confidences in the hypotheses that are relevant to the data at hand and that are as powerful as the particular application allows. The resulting assessments for simple hypotheses take the form of a minimum and a maximum confidence in each hypothesis.

## 2. CONFIDENCE IN HYPOTHESES

It is convenient to introduce some notation and definitions at this point, following as much as possible the notation and definitions in Bornholt (2025). We assume that for each possible value of a parameter $\theta$, the real valued vector random variable $X = X_1, X_2, \ldots, X_n$, has observed value $x$, has sample space $S \subset \mathbb{R}^n$, and the parameter space of $\theta$ is denoted $\Omega$. Either the random variable $X$ has an absolutely continuous distribution function with density $f(x;\theta)$ or it has a discrete distribution with frequency function $f(x;\theta)$. We assume the sample space $S$ is minimal in the sense that $x \in S$ implies $f(x;\theta) > 0$ for some $\theta$; $X$ is possibly a sufficient statistic from a more basic model; and in the continuous case that $f(x;\theta)$ denotes some fixed choice from the equivalent (almost everywhere) forms of the densities. "Model", "experiment" and "application" are alternative words used to denote $(X, f(x;\theta), \Omega)$.

Let $\theta_0$ and $\theta_1$ denote the two possible values of $\theta$ that determine the competing simple hypotheses, $H_0: \theta = \theta_0$ and $H_1: \theta = \theta_1$. Define the parameter $H = H(\theta)$ by $H = H_i$ iff $\theta = \theta_i$ for $i$ =0, 1. Thus, by assumption one value of $H$ is the true hypothesis. Confidences in the hypotheses are constructed from conditional probabilities that the maximum likelihood hypothesis and the minimum likelihood hypothesis estimators select the correct hypothesis. Specifically, $\widehat{H} = \widehat{H}(X)$ denotes the maximum likelihood estimator and $\widetilde{H} = \widetilde{H}(X)$ denotes the related minimum likelihood hypothesis estimator. Specifically,

$$\widehat{H} = \begin{cases} H_0 & (\text{and } \widetilde{H}{=}H_1) \quad \text{when } f(X;\theta_0) > f(X;\theta_1) \\ H_1 & (\text{and } \widetilde{H}{=}H_0) \quad \text{otherwise.} \end{cases} \tag{1}$$

Let $\hat{h} = \widehat{H}(x)$ denote the observed value of $\widehat{H}$, $\tilde{h} = \widetilde{H}(x)$ denote the observed value of $\widetilde{H}$, and let the expression "$\widehat{H}$ true" mean that $\widehat{H}$ selects the true hypothesis. Note that $\widehat{H}(\widetilde{H})$ selects the correct hypothesis when $\theta = \theta_i$ if $\widehat{H}(\widetilde{H}) = H_i$.

*Remark 1.* As it stands, equation (1) means $\widehat{H} = H_1$ when $f(X;\theta_0) = f(X;\theta_1)$. This aspect of the definition is arbitrary, but could only be an issue if $P_\theta\{f(X;\theta_0) = f(X;\theta_1)\} > 0$ for either $\theta$.

Determining the relevant conditional probabilities involves choosing a partition of the sample space prior to observing $X$. Let $\boldsymbol{C} = \{C_b, b \in B\}$ denote a partition of the sample space, where $b$ is a label from a set of labels $B$ and each disjoint subset $C_b$ is called a component of $\boldsymbol{C}$. (Each partition $\boldsymbol{C}$ implicitly defines a conditioning variable, say $Z^*$, having range $B$, i.e., $Z^* = b$ if and only if $X \in C_b$.) Different partitions are identified by bolded superscripts on $\boldsymbol{C}$, e.g. $\boldsymbol{C}^{\mathbf{1}}$ or $\boldsymbol{C}^{\mathbf{2}}$ etc. The subscripts $_0$, $_1$, and $_\theta$ in the notation $P_0, P_1$, and $P_\theta$ (and also $E_0[\,], E_1[\,]$, and $E_\theta[\,]$) denote probabilities or expectations under $\theta_0, \theta_1$, and $\theta$, respectively.

Let $\Gamma_{b\theta}$ denote the conditional probability that $\widehat{H}$ selects the correct hypothesis given $x \in C_b$. It follows that $1 - \Gamma_{b\theta}$ is the corresponding conditional probability that $\widetilde{H}$ selects the correct hypothesis given $x \in C_b$. That is,

$$\Gamma_{b\theta} = P_\theta\big(\hat{H} \text{ true} | X \in C_b\big) = P_\theta\big(\hat{H} = H | X \in C_b\big), \text{ and thus}$$
$$1 - \Gamma_{b\theta} = P_\theta\big(\tilde{H} \text{ true} \big| X \in C_b\big) = P_\theta\big(\tilde{H} = H \big| X \in C_b\big). \tag{2}$$

Now $\Gamma_{b\theta}$ has two possible values for each *b*: $\Gamma_{b0} = P_0\big(\hat{H} = H_0 | X \in C_b\big)$ and $\Gamma_{b1} = P_1\big(\hat{H} = H_1 | X \in C_b\big)$. The dependence of (2) on $\theta$ means that in general there will be a minimum and a maximum confidence in $\hat{h}$ and in $\tilde{h}$. While it is the minimum confidence in $\hat{h}$ (or equivalently the maximum confidence in $\tilde{h}$) that is of primary interest, the maximum confidence in $\hat{h}$ is retained because the assertion of, say, at least 70% and at most 75% confidence in $\hat{h}$ is considered more informative than the simpler assertion of at least 70% confidence in $\hat{h}$.

Accordingly, we use a closed interval $\text{Conf}(\hat{h}) = [\alpha_L(x), \alpha_U(x)]$ to represent confidence in $\hat{h}$, where $0 \leq \alpha_L(x) \leq \alpha_U(x) \leq 1$. The level of confidence in $\tilde{h}$ is the complementary interval $\text{Conf}(\tilde{h}) = [1 - \alpha_U(x), 1 - \alpha_L(x)]$. It is important to emphasise that intervals are being used to reflect the level of confidence not because confidence is based on some exotic conception of probability as an interval, but simply because the conditional probability $\Gamma_{b\theta}$ may depend on $\theta$ and we don't know which of the two possible values of $\theta$ is correct. Only the endpoints of the intervals are relevant to the confidence claims.

*Remark 2.* At this stage, we have not shown that if $\text{Conf}(\hat{h}) = [\alpha_L(x), \alpha_U(x)]$ then the level of confidence in $\tilde{h}$ must be the complementary interval $\text{Conf}(\tilde{h}) = [1 - \alpha_U(x), 1 - \alpha_L(x)]$. We postpone the justification for this result until Section 4.2, by which time we will have more-fully defined $\alpha_L(x)$ and $\alpha_U(x)$. The focus on confidence in $\hat{h}$ in what follows until that point is simply a matter of convenience. It does not signify that $\hat{h}$ is being singled out for special treatment nor is it being selected in some decision sense. Like the confidence interval method, hypotheses assessment is purely an inference procedure.

To guarantee Conf($\hat{h}$) is a valid frequentist confidence for partition $\boldsymbol{C}$ requires that

$$\alpha_L(x) \leq \Gamma_{b\theta} \leq \alpha_U(x) \tag{3}$$

for both $\theta$, all $x \in C_b$, and all $b \in B$. If (3) holds and $X = x$, we may state "at least $100\alpha_L(x)$% and at most $100\alpha_U(x)$% confidence in $\hat{h}$". (Of course, the simpler claim of "at least $100\alpha_L(x)$% confidence in $\hat{h}$" is also acceptable.)

We illustrate why assessments need to be conditional with a very-simple discrete example.

*Example 1.* Let Table 1 define the example. Now

$$\begin{aligned} P_0(\hat{H}\text{ true}) &= P_0(\hat{H} = H_0) = P_0(X = x_1 \text{ or } x_2) = 0.82, \text{ and} \\ P_1(\hat{H}\text{ true}) &= P_1(\hat{H} = H_1) = P_1(X = x_3 \text{ or } x_4) = 0.80, \end{aligned} \tag{4}$$

which means that it would be possible to claim at least 80% at most 82% confidence in $\hat{h}$. The difficulty with this claim is that intuitively we know from the respective likelihood ratios that $\hat{h}$ is more plausible when $X = x_1$ or $x_4$ than when $X = x_2$ or $x_3$. Although confidence levels based on (4) for all $x$ are probability-based, they nevertheless seem misleading *as measures of the relative plausibility of* $\hat{h}$ because we are only in a weak average sense "at least 80% and at most 82% confident in $\hat{h}$". In short, the claims are not relevant to the data at hand.

To see this, consider the partition $\boldsymbol{C}^1$ having two components $C_1 = \{x_1, x_4\}$ and $C_2 = \{x_2, x_3\}$ yielding

$$\left.\begin{aligned} \Gamma_{1\theta} &= 0.98 \quad \text{for both } \theta, \\ \Gamma_{20} &= 0.66 \quad \text{and} \\ \Gamma_{21} &= 0.62. \end{aligned}\right\} \tag{5}$$

Intuitively, it seems more sensible to state 98% confidence in $\hat{h}$ when $x \in C_1$ and at least 62% and at most 66% confidence in $\hat{h}$ when $x \in C_2$, rather than to state at least 80% and at most 82% confidence in $\hat{h}$ for all $x$. In particular, stating at least 80% confidence in $\hat{h}$ for any $x \in C_2$ seems a misleadingly-high claim when we know $P_\theta(\hat{H}\text{ true}|X \in C_2) \leq 0.66$. With partition $\boldsymbol{C}^1$ selected, if $X = x_2$ is observed then $\hat{h} = H_0$ and we can state at least 62% and at most 66% confidence in $H_0$ and at least 34% and at most 38% confidence in $H_1$.

**TABLE 1** Probabilities of various *x*

| $x$ | $f(x;\theta_0)$ | $f(x;\theta_1)$ | $\hat{h}$ |
|---|---|---|---|
| $x_1$ | 0.49 | 0.01 | $H_0$ |
| $x_2$ | 0.33 | 0.19 | $H_0$ |
| $x_3$ | 0.17 | 0.31 | $H_1$ |
| $x_4$ | 0.01 | 0.49 | $H_1$ |

*Remark 3.* Of course, conditional confidence procedures are not immune from having poor conditional properties. Returning to Example 1, suppose confidence claims had been based on a different partition $\boldsymbol{C}^2 = \{C_3, C_4\}$, where $C_3 = \{x_1, x_3\}$ and $C_4 = \{x_2, x_4\}$, In this case $P_\theta(\hat{H}\text{ true}|X \in C_3) =$ 0.742 or 0.969, and $P_\theta(\hat{H}\text{ true}|X \in C_4) =$ 0.721 or 0.971. It seems unreasonable to claim at least 72.1% confidence in $\hat{h}$ when $X = x_2$ and to claim at least 74.2% confidence in $\hat{h}$ when $X = x_3$, given that $C_2 = \{x_2, x_3\}$ and we know in advance that $P_\theta(\hat{H}\text{ true}|X \in C_2) \leq 0.66$.

## 3. AVOIDING MISLEADING CONFIDENCE CLAIMS

As Example 1 above makes clear, confidence levels satisfying (3) may be insufficiently relevant to the data at hand. Rather than relying on intuitive arguments as was done in that example, we need to consider applying one of the conditionality principles that have been proposed in the literature as ways of ensuring acceptable conditional properties. There are two basic types of conditionality principles, those that specify what to condition on if certain requirements are met, and those that require avoidance of procedures with, in some sense, poor conditional properties. It is the latter type that is relevant for our purpose.

In a landmark paper, Robinson (1979) provided a systematic set of definitions of desirable conditional properties. One was a *squared-error loss* admissibility criterion and the others required the absence of one of three types of betting strategies called *semirelevant, relevant* and *super-relevant* strategies or procedures. The nonexistence of semirelevant strategies is a stronger condition than the nonexistence of relevant strategies, and the nonexistence of relevant strategies is a stronger condition than the nonexistence of super-relevant strategies. Squared-error loss admissibility is a property intermediate between the absence of semirelevant and relevant betting procedures. As a result, the absence of relevant and super-relevant betting procedures can be regarded as two types of extended or weak squared-error loss admissibility. This gives the absence of these betting procedures a non-betting inference interpretation, perhaps making the imposition of either of these conditions more acceptable to those readers who might question the relevance of betting to the statistician's aims.

Bondar (1977) and Robinson (1979) both argued that the absence of relevant *subsets* is too stringent a requirement because it would eliminate the usual Student's *t* confidence interval. Since the absence of relevant subsets is a weaker requirement than the absence of relevant betting procedures, their argument implies that the absence of relevant betting procedures is also too stringent. Thus, for hypotheses assessments we choose the weakest of Robinson's four conditions (absence of super-relevant betting procedures) as our consistency requirement because this requirement is still strong enough to eliminate severe examples of poor conditional properties such as those discussed for Example 1. Note that our consistency condition is marginally stronger than Bondar's (1977) consistency condition (absence of super-relevant *subsets*).

### *3.1. Consistency and super-relevant betting procedures.*

We need to extend Robinson's (1979) definition of super-relevant betting procedures in order to allow for the interval nature of $[\alpha_L(x), \alpha_U(x)]$. Consider a hypothetical betting game between two players, Peter who proposes $[\alpha_L(x), \alpha_U(x)]$ as his confidence in $\hat{h}$ and Paula who questions the reasonableness of $[\alpha_L(x), \alpha_U(x)]$ for some $x$ and is willing to bet against it. Treating $[\alpha_L(x), \alpha_U(x)]$ like a probability assertion, Peter is required to offer odds $\{1 - \alpha_U(x)\}: \alpha_U(x)$ should Paula wish to bet $\hat{h}$ is true, and to offer odds $\alpha_L(x): \{1 - \alpha_L(x)\}$ should Paula wish to bet $\hat{h}$ is false. Paula bets against $[\alpha_L(x), \alpha_U(x)]$ with strategy denoted by $(u(x), s(x))$, where $u(x) \in \{0,1\}$ and $0 \leq s(x) \leq 1$ and where; (i) when $u(x) = 0$ and $s(x) > 0$, Paula places a bet of size $\alpha_U(x)s(x)$ that $\hat{h}$ is the true hypothesis; (ii) when $u(x) = 1$ and $s(x) > 0$, she places a bet of size $(1 - \alpha_L(x))s(x)$ that $\hat{h}$ is false; and (iii) when $s(x) = 0$, no bet is made and $u(x)$ is set to zero. For those $x$ such that $s(x) > 0$, the function $u(x)$ is the indicator function of Paula betting $\hat{h}$ is false.

The pair $(u(x), s(x))$ is called a betting procedure or betting strategy, and is assumed to be nontrivial in the sense that $E_\theta[s(X)] > 0$ for some $\theta$. Note that Robinson's $|s(x)|$ is our $s(x)$, his $s(x)$ is our $\{1 - 2u(x)\}s(x)$. For definiteness, we chose to bound $s(x)$ above by 1 because we are interested in whether or not certain types of winning betting procedures exist, and the answer is unaffected by whatever nonzero upper bound on $s(x)$ we choose. Peter makes known his whole function $[\alpha_L(x), \alpha_U(x)]$, and Paula announces her betting strategy $(u(x), s(x))$ before the game commences. A referee selects $\theta$, $X = x$ is observed, Paula's bet if any is placed according to the announced strategy at the required odds, $\theta$ and hence $H$ are revealed, and the game settled. Define the indicator function $\mathbf{1}_{\hat{h}}(H) = 1$ if $\hat{h} = H$, and otherwise $\mathbf{1}_{\hat{h}}(H) = 0$. Paula's expected return can be written

$$R_\theta = E_\theta\Big[\Big\{-\big(\mathbf{1}_{\hat{H}}(H) - \alpha_L(X)\big)u(X) + \big(\mathbf{1}_{\hat{H}}(H) - \alpha_U(X)\big)\big(1 - u(X)\big)\Big\}s(X)\Big]\,.$$

Then the betting procedure $(u(x), s(x))$ is said to be *super-relevant* if for some $\varepsilon > 0$, $R_\theta \geq \varepsilon$ for both $\theta$. Thus, the existence of a super-relevant betting procedure means that there exists a betting strategy with positive expected return bounded away from zero.

The absence of super-relevant betting procedures is imposed on $[\alpha_L(x), \alpha_U(x)]$ as follows. For $k = 0, 1$ and arbitrary scalars $\pi_k$, where $0 \leq \pi_0 \leq 1$ and $\pi_1 = 1 - \pi_0$, consider the equation

$$\sum_k \{\mathbf{1}_{\hat{h}}(H_k) - q(x, \pi_0)\} f(x; \theta_k)\pi_k = 0 \quad \text{for all } x. \tag{6}$$

Unless $\sum_k f(x; \theta_k)\pi_k = 0$, equation (6) defines the function $q(x, \pi_0)$ as

$$q(x, \pi_0) = \sum_k \mathbf{1}_{\hat{h}}(H_k) f(x; \theta_k)\pi_k \,/\, \sum_k f(x; \theta_k)\pi_k. \tag{7}$$

That is, $q(x, \pi_0)$ is defined for all $x$ for which $\sum_k f(x; \theta_k)\pi_k \neq 0$. Then a sufficient condition for the nonexistence of super-relevant betting procedures is implied by the following theorem.

*Theorem 1. Exactly one of the following alternatives holds: (i) there exists a betting procedure* $(u(x), s(x))$ *for* $[\alpha_L(x), \alpha_U(x)]$ *such that* $R_\theta > 0$ *for both θ, or (ii) there exists some* $\pi_0$ *such that*

$$\alpha_L(x) \leq q(x, \pi_0) \leq \alpha_U(x), \qquad \textit{a.e.} \tag{8}$$

*(with respect to the probability induced by* $f(x; \theta)$*) for those x for which* $q(x, \pi_0)$ *is defined.*

*Proof.* The proof can be found in the Appendix.

Imposing (8) for any $\pi_0$ rules out betting strategies with $R_\theta > 0$ for both $\theta$, and hence ensures the non-existence of super-relevant betting strategies (those with $R_\theta \geq \varepsilon$ for both $\theta$). Henceforth $[\alpha_L(x), \alpha_U(x)]$ will be called *consistent* if it is free of super-relevant betting procedures. Consistency will be guaranteed by imposing (8) for any $\pi_0$.

*Remark 4.* The confidence levels $\alpha_L(x)$ and $\alpha_U(x)$ bracket $q(x, \pi_0)$ which has the same value as would the posterior probability of $\hat{h}$ were a Bayesian to use $\pi_0$ as the prior probability of $H_0$. However, in hypotheses assessments $\pi_0$ is just a proportion determined internally by the admissibility criteria described in the following section.

## 4. THE ASSESSMENT METHOD

Recall that our aim is to determine frequentist confidences in the hypotheses that are relevant to the data at hand and are as powerful as the particular circumstances allow. Confidences being probability-based led to constraint (3) given a partition $\boldsymbol{C}$, while the use of confidences as relevant measures of relative plausibility led to the consistency constraint (8) for any $\pi_0$ satisfying $0 \leq \pi_0 \leq 1$. Accordingly, given a particular choice $(\boldsymbol{C}, \pi_0)$ made prior to observing an $x$ and given the desire not to make the intervals of confidence any wider than necessary, we incorporate both constraints by defining $[\alpha_L(x), \alpha_U(x)]$ by

$$\begin{aligned} \alpha_L(x) &= \inf\{\Gamma_{b0}, \Gamma_{b1}, q(x, \pi_0)\} \\ \alpha_U(x) &= \sup\{\Gamma_{b0}, \Gamma_{b1}, q(x, \pi_0)\} \end{aligned} \tag{9}$$

for all $x \in C_b$, all $b \in B$, and this $\pi_0$. In effect, the assessment method makes an adjustment to ensure consistency for any $x$ for which either $q(x, \pi_0) < \inf\{\Gamma_{b0}, \Gamma_{b1}\}$ or $q(x, \pi_0) > \sup\{\Gamma_{b0}, \Gamma_{b1}\}$. The net effect of any such consistency adjustments would be that the resulting confidence claims will be more-conservative than had there been no adjustments for consistency.

### *4.1. Admissible hypotheses assessments.*

For every pair $(\boldsymbol{C}, \pi_0)$, we use the notation $[\alpha_L(x, \boldsymbol{C}, \pi_0), \alpha_U(x, \boldsymbol{C}, \pi_0)]$ when wishing to indicate the dependence of $[\alpha_L(x), \alpha_U(x)]$ on the pair $(\boldsymbol{C}, \pi_0)$. To choose an appropriate $(\boldsymbol{C}, \pi_0)$ we need a performance criterion, a notion of goodness. Since it is the minimum confidence in $\hat{h}$ that matters the most, it seems natural to want expected minimum confidence in the maximum likelihood hypothesis estimator to be as large as possible. This suggests that the admissibility of a specific pair, $(\boldsymbol{C}^{\mathbf{1}}, \pi_{01})$ say, should depend on expected minimum confidence in the following way. The pair $(\boldsymbol{C}^{\mathbf{1}}, \pi_{01})$ is regarded as better than $(\boldsymbol{C}^{\mathbf{2}}, \pi_{02})$ (first sense) if

$$E_\theta[\alpha_L(X, \boldsymbol{C}^{\mathbf{2}}, \pi_{02})] \leq E_\theta[\alpha_L(X, \boldsymbol{C}^{\mathbf{1}}, \pi_{01})]$$

for both $\theta$ with strict inequality for some $\theta$. The pair $(\boldsymbol{C^1}, \pi_{01})$, and hence

$$[\alpha_L(x), \alpha_U(x)] = [\alpha_L(x, \boldsymbol{C^1}, \pi_{01}), \alpha_U(x, \boldsymbol{C^1}, \pi_{01})]$$

are called <u>admissible (first sense)</u> if no better pair (first sense) exists.

As noted earlier, it is the minimum confidence in $\hat{h}$ that is of primary interest. If there is more than one admissible (first sense) pair then some of these pairs may lead to narrower intervals of confidence on average, and intervals narrower on average seem marginally preferable. This suggests a refinement of the above admissibility criterion. If both $(\boldsymbol{C^1}, \pi_{01})$ and $(\boldsymbol{C^2}, \pi_{02})$ are admissible (first sense) then $(\boldsymbol{C^1}, \pi_{01})$ is better than $(\boldsymbol{C^2}, \pi_{02})$ (second sense) if

$$E_\theta[\alpha_U(X, \boldsymbol{C^2}, \pi_{02}) - \alpha_L(X, \boldsymbol{C^2}, \pi_{02})] \\ \geq E_\theta[\alpha_U(X, \boldsymbol{C^1}, \pi_{01}) - \alpha_L(X, \boldsymbol{C^1}, \pi_{01})] \quad (10)$$

for both $\theta$ with strict inequality for some $\theta$. The pair $(\boldsymbol{C^1}, \pi_{01})$, and hence

$$[\alpha_L(x), \alpha_U(x)] = [\alpha_L(x, \boldsymbol{C^1}, \pi_{01}), \alpha_U(x, \boldsymbol{C^1}, \pi_{01})]$$

are called <u>admissible (second sense)</u> if they are admissible (first sense) and if no better pair (second sense) exists.

For all $(\boldsymbol{C}, \pi_0)$, (9) implies that

$$E_\theta[\alpha_L(X, \boldsymbol{C}, \pi_0)] \leq P_\theta(\widehat{H} = H) \quad \text{for both } \theta \quad (11)$$

because $E_\theta[\alpha_L(X, \boldsymbol{C}, \pi_0)] \leq E_\theta[\Gamma_{b\theta}] = P_\theta(\widehat{H} = H)$ for both $\theta$.

### *4.2. Admissibility of Conf*$(\tilde{h})$

As discussed in Remark 2, we need to justify setting $\text{Conf}(\tilde{h}) = [1 - \alpha_U(x), 1 - \alpha_L(x)]$. We consider in turn confidence, consistency and admissibility. Firstly, (9) implies $1 - \alpha_U(x) \leq 1 - \Gamma_{b\theta} \leq 1 - \alpha_L(x)$ for both $\theta$, all $x \in C_b$, and all $b \in B$, which means that $\text{Conf}(\tilde{h})$ is a valid confidence, given $1 - \Gamma_{b\theta} = P_\theta\left(\widetilde{H} \text{ true} \middle| X \in C_b\right)$ from (2).

Secondly, in the hypothetical betting game in Section 3.1, Peter is required to offer odds

$\{1-\alpha_U(x)\}$: $\alpha_U(x)$ should Paula wish to bet $\hat{h}$ is true, and to offer odds $\alpha_L(x)$: $\{1-\alpha_L(x)\}$ should Paula wish to bet $\hat{h}$ is false. Now in a "two-horse race", betting $\hat{h}$ is true is the same as betting $\tilde{h}$ is false, so Peter needs to offer the same odds for these equivalent events, $\{1-\alpha_U(x)\}$: $\alpha_U(x)$. Similarly, betting $\hat{h}$ is false is the same as betting $\tilde{h}$ is true so Peter needs to offer the same odds for these equivalent events, $\alpha_L(x)$: $\{1-\alpha_L(x)\}$.

We are now in a position to modify Theorem 1 for betting for or against $\tilde{h}$ being true. $\text{Conf}(\tilde{h})$ will be consistent if

$$1-\alpha_U(x) \leq q^*(x,\pi_0) \leq 1-\alpha_L(x) \qquad \text{a.e.,} \tag{12}$$

where, from (7), $q^*(x,\pi_0)$ is determined by modifying $q(x,\pi_0)$ by replacing $\mathbf{1}_{\hat{h}}(H_k)$ in $q(x,\pi_0)$ with $\mathbf{1}_{\tilde{h}}(H_k)$. That is,

$$q^*(x,\pi_0) = \sum\nolimits_k \mathbf{1}_{\tilde{h}}(H_k) f(x;\theta_k)\pi_k \,/ \sum\nolimits_k f(x;\theta_k)\pi_k.$$

It follows that $q^*(x,\pi_0) = 1-q(x,\pi_0)$ since $\mathbf{1}_{\tilde{h}}(H_k) = 1-\mathbf{1}_{\hat{h}}(H_k)$. The consistency condition can thus be rewritten as $1-\alpha_U(x) \leq 1-q(x,\pi_0) \leq 1-\alpha_L(x)$. This condition is implied by (9), meaning that the consistency of $\text{Conf}(\tilde{h})$ is automatically guaranteed whenever $\text{Conf}(\hat{h})$ is consistent (and vice versa).

Lastly, we come to the question of whether or not $\text{Conf}(\widetilde{H}) = [1-\alpha_U(X), 1-\alpha_L(X)]$ is admissible whenever $\text{Conf}(\widehat{H}) = [\alpha_L(X), \alpha_U(X)]$ is admissible. In Section 4.1, first-sense admissibility was based on the natural preference for the expected minimum confidence $E_\theta[\alpha_L(X)]$ in the maximum likelihood hypothesis estimator to be as large as possible. The corresponding preference for the minimum likelihood hypothesis estimator is for the expected maximum confidence $E_\theta[1-\alpha_L(X)]$ to be as small as possible. Clearly, making $E_\theta[1-\alpha_L(X)]$ as small as possible is the same as making $E_\theta[\alpha_L(X)]$ as large as possible, and thus $\text{Conf}(\widetilde{H}) = [1-\alpha_U(X), 1-\alpha_L(X)]$ is first-sense admissible whenever $\text{Conf}(\widehat{H}) = [\alpha_L(X), \alpha_U(X)]$ is first-sense admissible (and vice-versa).

Second-sense admissibility is similarly guaranteed for $\mathrm{Conf}(\widetilde{H}) = [1 - \alpha_U(X), 1 - \alpha_L(X)]$ whenever $\mathrm{Conf}(\widehat{H}) = [\alpha_L(X), \alpha_U(X)]$ is second-sense admissible. In Section 4.1, second-sense admissibility was based on the preference for narrower intervals between the upper and lower confidences on average if first-sense admissible. The interval range for $\mathrm{Conf}(\widetilde{H})$ is $1 - \alpha_L(X) - \big(1 - \alpha_U(X)\big) = \alpha_U(X) - \alpha_L(X)$. This is the same as the interval range of $\mathrm{Conf}(\widehat{H})$. Hence, $\mathrm{Conf}(\widetilde{H})$ is second-sense admissible whenever $\mathrm{Conf}(\widehat{H})$ is second-sense admissible (and vice-versa).

#### *4.3. Symmetric experiments.*

The upper bounds in (11) are achievable for symmetric experiments. Following Birnbaum (1961), an experiment is called symmetric if the likelihood ratio $\lambda(X) = f(X;\theta_1)/f(X;\theta_0)$ has the same distribution under $H_0$ as $1/\lambda(X)$ has under $H_1$. Symmetric applications arise, for example, if $f(x;\theta_0) = f(w - x;\theta_1)$ for all $x$ and some scalar $w$.

Let $v(x)$ denote the likelihood ratio in favour of $\hat{h}$. That is,

$$v(x) = f(x;\hat{\theta})/f(x;\tilde{\theta}) \quad \text{for all } x \in S. \tag{13}$$

Consider partitioning so that observations with the same $v(x)$ are all in the same component. For example, define the partition $\boldsymbol{C} = \{C_b, b \in B\}$ where $C_b = \{x: v(x)/[1 + v(x)] = b\}$ for all $b \geq$ ½ for which $v(x)$ exists. If, for any $x \in S$, equation (13) is undefined for some $x$ because $f\big(x;\tilde{h}\big) = 0$, lump such $x$ into the one component denoted $C_\infty$. For $x \notin C_\infty$, symmetry ensures that

$$\Gamma_{b0} = \Gamma_{b1} = \frac{v(x)}{[1 + v(x)]} \tag{14}$$

for all $x \in C_b$ and all $b \in B$. Moreover, if $\pi_0 =$ ½ is chosen then from (7)

$$q(x,½) = \frac{v(x)}{[1 + v(x)]} \qquad \text{for all } x \notin C_\infty. \tag{15}$$

Substituting (14) and (15) into (9) gives

$$\mathrm{Conf}(\hat{h}) = \alpha_L(x) = \alpha_U(x) = \frac{v(x)}{[1 + v(x)]} \qquad \text{for all } x \notin C_\infty. \tag{16}$$

For all $x \in C_\infty$, note that $f(x; \tilde{h}) = 0$ means that $P_\theta(\hat{H} \text{ true} | X \in C_\infty) = 1 = q(x,½)$, leading to $\alpha_L(x) = \alpha_U(x) = 1$ from (9), and hence Conf($\hat{h}$) = 1 for all $x \in C_\infty$.

It is easy to see that the upper bounds in (11) are reached for both $\theta$. Also, the right-hand side of (10) is zero for both $\theta$ for this choice of partition and $\pi_0$. Thus this ($\boldsymbol{C}$,½) is optimal (second sense), as are the resulting confidence claims. A feature of symmetric experiments is that no adjustments are needed to achieve consistency. Translating Conf($\hat{h}$) back to confidences in the original hypotheses using $\lambda(x) = f(x; \theta_1)/f(x; \theta_0)$ yields $\mathrm{Conf}(H_0) = 1/\{1 + \lambda(x)\}$ and $\mathrm{Conf}(H_1) = \lambda(x)/\{1 + \lambda(x)\}$. Thus, in symmetric experiments our confidence in $H_i$ has the same *value* as the Bayesian posterior probability of $H_i$ that would result from assuming equal prior probabilities for the two hypotheses.

*Example 2.* Suppose $X \sim N(\mu, \sigma^2 \mathrm{I}_n)$ with $\sigma^2$ known so that $\theta = \mu$, and the hypotheses to assess are $H_0$: $\mu = \mu_0$ and $H_1$: $\mu = \mu_1$, where $\mu_0$ and $\mu_1$ are known and $\mu_1 > \mu_0$. For this symmetric example, the optimal partition can also be written $\boldsymbol{C} = \{C_b, b \in B\}$, where

$$C_b = \{x : \bar{x} = b \text{ or } \mu_0 + \mu_1 - b\} \tag{17}$$

for all $b \geq (\mu_0 + \mu_1)/2$. Note that in this partition, $C_b$ for $b > (\mu_0 + \mu_1)/2$ contains just two $\bar{x}$ points, and is an example of what Kiefer (1977) called a fine continuum partition. The likelihood ratio $v(x) = \exp(\gamma|\gamma/2 - z|)$ for this example, where $\gamma = (\mu_1 - \mu_0)\sqrt{n}/\sigma$, and $z = (\bar{x} - \mu_0)\sqrt{n}/\sigma$. Confidence in $\hat{h}$ follows from substituting this $v(x)$ into (16). Thus, for example, if $\mu_0 = 0$, $\mu_1 = 2$, and $\sigma/\sqrt{n} = 1$ then $\bar{x} = 3$ would give confidence in $H_0$ of 1.8% and confidence in $H_1$ of 98.2%.

*Remark 5*. For the special case of Example 2 with $\mu_1 = -\mu_0$, the competing hypotheses are $H_0: \mu = -\mu_1$ versus $H_1: \mu = \mu_1$, where $\mu_1 > 0$ is known. The optimal partition from (17) simplifies to $C_b = \{x: \bar{x} = b \text{ or } -b\}$ for all $b \geq 0$, and is same partition irrespective of the particular value of $\mu_1 (> 0)$. The likelihood ratio $v(x)$ simplifies to $v(x) = \exp(2n\mu_1|\bar{x}|/\sigma^2) = \exp(2n|\mu||\bar{x}|/\sigma^2)$. Interestingly, except for when $b = 0$, this partition and likelihood ratio are identical to the corresponding items in Equations (7) and (8) in Bornholt (2025) for the one-sided composite hypotheses case with $H_0: \mu \leq 0$ versus $H_1: \mu > 0$. This link seems appropriate because this set of one-sided composite hypotheses can also be described as $H_0: \mu = -\mu_1 (or\ 0)$ versus $H_1: \mu = \mu_1$, where $\mu_1 > 0$ is *unknown*. Overall, this connection between Example 2 and the one-sided hypotheses case provides further support for the partition used in Bornholt (2025).

### *4.4. Asymmetric experiments.*

The agreement in value, though not in interpretation, between Conf($H_i$) and the Bayesian measure for the relative plausibility of $H_i$ does not carry over to asymmetric experiments. We illustrate hypotheses assessment for the asymmetric case with a simple genetics model.

*Example 3.* The competing hypotheses are $H_0$: $\theta_0 = 0.25$ and $H_1$: $\theta_1 = 0.5$, where $\theta$ is the probability that each of certain progeny have a particular trait. Let $X = x$ be the number of progeny observed to have the trait, where $X$ is a binomial random variable based on 10 independent Bernoulli trials. Columns 2 and 3 of Table 2 provide the (rounded) probabilities involved.

A reasonable way to proceed in asymmetric cases is to consider partitions in which each component $C_b$ has the property that $\hat{h} = H_0$ for at least one $x$ and $\hat{h} = H_1$ for at least one other $x$. Moreover, the members of a particular component should have similar values of $v(x)$, to the extent allowed by the asymmetry of the application (recall that the $x$ values in each optimal component $C_b$ for *symmetric* experiments all have the same values for $v(x)$). Inspection of the $v(x)$ likelihood ratios in Table 2 suggests as an obvious first choice the partition $\boldsymbol{C^1}$ with components $C_1$ = {3,4}, $C_2$ = {2,5}, $C_3$ = {1,6} and $C_4$ = {0,7,8,9,10}. This partition yields

$$\Gamma_{10}, \Gamma_{11} = 0.632, 0.636; \Gamma_{30}, \Gamma_{31} = 0.920, 0.955$$
$$\Gamma_{20}, \Gamma_{21} = 0.828, 0.848; \Gamma_{40}, \Gamma_{41} = 0.941, 0.994$$

A search of other likely alternative partitions suggests that $\boldsymbol{C^1}$ is the only partition that is admissible (first sense). Regarding consistency adjustments with $\boldsymbol{C^1}$, any $\pi_0$ in the range $0.446 \leq \pi_0 \leq 0.45$ was found to be admissible (first sense), while $\pi_0 = 0.45$ was the optimal value (second sense). The final three columns of Table 2 list the various values for *b*, *q*(*x*,0.45), and Conf($\hat{h}$), respectively, for $(\boldsymbol{C^1}, 0.45)$. Thus, for example, if *x* = 6 is observed then we state at least 4.5% and at most 8% confidence in $H_0$ and at least 92% and at most 95.5% confidence in $H_1$. [The *p*-value for *x* = 6 is 0.0197.]

**TABLE 2** Summary of Example 3 based on $(\boldsymbol{C^1}, 0.45)$

| $x$ | $f(x;\theta_0)$ | $f(x;\theta_1)$ | $\hat{h}$ | $v(x)$ | $b$ | $q(x,0.45)$ | Conf($\hat{h}$) |
|---|---|---|---|---|---|---|---|
| 0 | $0.563\times10^{-1}$ | $0.977\times10^{-3}$ | $H_0$ | 57.4 | 4 | 0.979 | [0.941, 0.994] |
| 1 | 0.188 | $0.977\times10^{-2}$ | $H_0$ | 19.2 | 3 | 0.940 | [0.920, 0.955] |
| 2 | 0.282 | $0.439\times10^{-1}$ | $H_0$ | 6.4 | 2 | 0.840 | [0.828, 0.848] |
| 3 | 0.250 | 0.117 | $H_0$ | 2.1 | 1 | 0.636 | [0.632, 0.636] |
| 4 | 0.146 | 0.205 | $H_1$ | 1.4 | 1 | 0.632 | [0.632, 0.636] |
| 5 | $0.584\times10^{-1}$ | 0.246 | $H_1$ | 4.2 | 2 | 0.837 | [0.828, 0.848] |
| 6 | $0.162\times10^{-1}$ | 0.205 | $H_1$ | 12.6 | 3 | 0.939 | [0.920, 0.955] |
| 7 | $0.309\times10^{-2}$ | 0.117 | $H_1$ | 37.9 | 4 | 0.979 | [0.941, 0.994] |
| 8 | $0.386\times10^{-3}$ | $0.439\times10^{-1}$ | $H_1$ | 113.8 | 4 | 0.993 | [0.941, 0.994] |
| 9 | $0.286\times10^{-4}$ | $0.977\times10^{-2}$ | $H_1$ | 341.3 | 4 | 0.9976 | [0.941, 0.9976†] |
| 10 | $0.954\times10^{-6}$ | $0.977\times10^{-3}$ | $H_1$ | 1024 | 4 | 0.9992 | [0.941, 0.9992†] |

† These two values are larger due to a consistency adjustment.

This example includes an adjustment for consistency in the final column to the upper confidence limit for *x* = 9 and *x* = 10 to ensure *q*(*x*,0.45) is not above $\alpha_U(x)$ for these *x*'s. This adjustment seems intuitively reasonable as the observations in $C_4$ have a very wide range of values for *v*(*x*), from 37.9 to 1024. Consistency is indicating that a maximum confidence level for $\hat{h}$ of 99.4% is a little too low for the two observations with the largest *v*(*x*) values in $C_4$.

## 5. DISCUSSION

This article extends Bornholt’s (2025) hypotheses assessment method to the case with two simple hypotheses. Studying the simple hypotheses case provides clarification for a number of issues, including the relationships between assessments and testing, between different types of evidence, and between assessments and posterior probabilities.

## Appendix

### Proof of Theorem 1

**Proof.** Suppose *X* is non-discrete. Recall that Paula's expected return is

$$R_\theta = E_\theta\Big[\big\{-\big(\mathbf{1}_{\widehat{H}}(H) - \alpha_L(X)\big)u(X) + \big(\mathbf{1}_{\widehat{H}}(H) - \alpha_U(X)\big)\big(1-u(X)\big)\big\}s(X)\Big]\,.$$

Applying Lemma 2 of Buehler (1976), we know that exactly one of the following holds: (i') there exists bounded functions $s(x) \geq 0$ for all *x*, such that $R_\theta > 0$ for both $\theta$, or (ii') there exists scalars $w_k \geq 0$ for both *k* but not all zero, such that

$$\textstyle\sum_k \big\{-\big(\mathbf{1}_{\hat{h}}(H_k) - \alpha_L(x)\big)u(x) + \big(\mathbf{1}_{\hat{h}}(H_k) - \alpha_U(x)\big)\big(1-u(x)\big)\big\}f(x;\theta_k)w_k \leq 0 \quad \text{a.e.}$$

with respect to the probability induced by $f(x;\theta)$, or equivalently, such that

$$\textstyle\sum_k \big\{-\big(\mathbf{1}_{\hat{h}}(H_k) - \alpha_L(x)\big)u(x) + \big(\mathbf{1}_{\hat{h}}(H_k) - \alpha_U(x)\big)\big(1-u(x)\big)\big\}f(x;\theta_k)\pi_k \leq 0 \quad \text{a.e.} \tag{18}$$

with respect to the probability induced by $f(x;\theta)$, if we let $\pi_0 = w_0/(w_0+w_1)$ and $\pi_1 = 1-\pi_0$. Now for those *x* such that $\sum_k f(x;\theta_k)\pi_k = 0$, (18) holds. [Equation (18) holds because $\sum_k f(x;\theta_k)\pi_k = 0$ implies $f(x;\theta_k)\pi_k = 0$ for both *k* for these *x* due to the nonnegativity of the $f(x;\theta_k)$'s and of the $\pi_k$'s. Hence, the left-hand side of (18) equals zero for these *x*.] For each of the remaining *x*'s (for which $\sum_k f(x;\theta_k)\pi_k \neq 0$ and hence for which $q(x,\pi_0)$ is defined), inserting each of the two possible values for *u*(*x*) in turn into (18) reproduces equation (8) [*u*(*x*) = 0 yields $q(x,\pi_0) \leq \alpha_U(x)$ a.e., while *u*(*x*) = 1 yields $\alpha_L(x) \leq q(x,\pi_0)$ a.e.]. If *X* is discrete, the proof follows similarly by applying Lemma 1 of Buehler (1976). **||**